%% file: IEEEBigData2019.tex
\documentclass[conference]{IEEEtran}
\usepackage{caption}
\newtheorem{algorithm}{Algorithm}

\usepackage{algorithm}
\usepackage[noend]{algpseudocode}

\usepackage{varwidth}
\usepackage{amsmath}

\usepackage{hyperref}
\hypersetup{colorlinks=true,citecolor = blue}

\newtheorem{definition}{Definition} %gws
\usepackage{graphicx,epstopdf,algpseudocode,caption,url}   % gws
\usepackage{multirow}  %gws 201700930
\graphicspath{{./images}}
\usepackage{arydshln} % gws, 20180202

\usepackage{color}   % gws 2017-10-22
\usepackage{graphicx}   % gws 2017-10-22
\usepackage{graphicx,epstopdf,caption,url}   % gws
\usepackage{multirow}  %gws 20170930
\usepackage{graphicx,epstopdf,algpseudocode,caption,url}   % gws
\ifCLASSINFOpdf
\else
\fi

\begin{document}
%\title{A Survey of Privacy Preserving Utility Mining} %  A Survey and New Perspectives
\title{Privacy Preserving Utility Mining: A Survey}

\author{\IEEEauthorblockN{Wensheng Gan$ ^{1,5} $,
		Jerry Chun-Wei Lin$ ^{1,2*}$,
		Han-Chieh Chao$ ^{3} $,
		Shyue-Liang Wang$ ^{4} $ and
		Philip S. Yu$ ^{5} $}
	\IEEEauthorblockA{$ ^{1} $Harbin Institute of Technology (Shenzhen), Shenzhen, China} \IEEEauthorblockA{$ ^{2} $Western Norway University of Applied Sciences, Bergen, Norway}
 \IEEEauthorblockA{$ ^{3} $National Dong Hwa University, Hualien, Taiwan}
 	\IEEEauthorblockA{$ ^{4} $National University of Kaohsiung, Kaohsiung, Taiwan}
	\IEEEauthorblockA{$ ^{5} $University of Illinois at Chicago, Chicago, USA}
	\IEEEauthorblockA{
		Email: wsgan001@gmail.com, jerrylin@ieee.org, hcc@ndhu.edu.tw, slwang@nuk.edu.tw, psyu@uic.edu
	% \IEEEauthorblockA{\IEEEauthorrefmark{3}\\
	% Telephone: (800) 555--1212, Fax: (888) 555--1212}
}

}

\maketitle

\begin{abstract}
In big data era, the collected data  usually contains rich information and hidden knowledge. Utility-oriented pattern mining and analytics have shown a powerful ability to explore these ubiquitous data, which may be collected from various fields and applications, such as market basket analysis, retail, click-stream analysis, medical analysis, and bioinformatics. However, analysis of these data with sensitive private information raises privacy concerns. To achieve better trade-off between utility maximizing and privacy preserving, Privacy-Preserving Utility Mining (PPUM)  has become a critical issue in recent years. In this paper, we provide a comprehensive overview of PPUM. We first present the background of utility mining, privacy-preserving data mining and PPUM, then introduce the related preliminaries and problem formulation of PPUM, as well as some key evaluation criteria for PPUM. In particular, we present and discuss the current state-of-the-art PPUM algorithms, as well as their advantages and deficiencies in detail. Finally, we highlight and discuss some technical challenges and open directions for future research on PPUM.

\end{abstract}

% privacy, data mining, privacy-preserving data mining

% \fi
\IEEEpeerreviewmaketitle

\input{1_intro.tex}

\input{2_relatedwork.tex}

\input{3_preliminaries.tex}

\input{4_survey.tex}

\input{5_opportunities.tex}

\bibliographystyle{IEEEtran}
\bibliography{paper}

\end{document}

%% file: 1_intro.tex
\section{Introduction}

With the development of technologies and the pervasiveness of computing devices, various types of data dramatically  increase. The collected data in the big data era usually contains auxiliary information and hidden knowledge. Great advances in computing technologies bring many benefits to society, with transformative changes and financial opportunities being created in health-care, transportation, industry, education, commerce, and social interactions. During the whole process from data collection to knowledge discovery, the data  typically contains sensitive and individual information such as medical and financial information, and they may be exposed to several parties including collectors, owners, users and miners. However, the collected information may contain sensitive and private information, which raises \textit{privacy} concerns \cite{bertino2008survey}. Up to now, there is no standard definition of \textit{privacy} concept. Bertino et al. \cite{bertino2008survey}  gave a definition of privacy, in terms of the control of the data, but explicitly incorporate the risks of privacy violation.  

Data mining and analytics \cite{aggarwal2008general,han2004mining} have shown a powerful ability to explore data (usually large amounts of data, also known as ``big data''). Therefore, data mining technologies are commonly used in many real-world applications to extract hidden valuable knowledge by analyzing large amounts of data,  typically business data or other sensitive data. For the health-care data, these datasets contain confidential or secure information (i.e., personal identification number, social security number, credit card number, etc.) and lead to privacy threats if they are published in the public place or misused, especially for government agencies and commercial institutions. Data privacy and security are the key problems in data management and analytics. Transforming the data may reduce its utility, resulting in inaccurate or even infeasible extraction of knowledge through data mining. This is the paradigm known as Privacy-Preserving Data Mining (PPDM) \cite{aggarwal2008general,agrawal2000privacy,lindell2000privacy}. The common method to protect data is encryption, but it reduces the analyzability of data and is seldom used by data mining techniques. PPDM methodologies are designed to guarantee a certain level of privacy,  while maximizing the utility of data, such that data mining can still be performed on the transformed data efficiently.  PPDM conducts data mining operations under the condition of preserving data privacy. Most of them focus on two aspects, such as privacy and utility, and they have conflict relationship \cite{aggarwal2008general,agrawal2000privacy,lindell2000privacy}. With consideration of privacy and utility, PPDM aims at achieving better trade-off between utility maximizing and privacy preserving. Both privacy and utility are critically important for PPDM. Here the term of ``\textit{utility}'' can be referred as the \textit{availability} of data.

Broadly, the PPDM methodologies can be roughly divided into several categories \cite{bertino2008survey,aggarwal2008general}. Detailed algorithms and methodologies of each category can be referred to \cite{bertino2008survey,aggarwal2008general,dwork2008differential}. Aggarwal et al. \cite{aggarwal2008general} presented a detailed survey on some techniques used for PPDM. Broadly, based on Aldeen et al. \cite{aldeen2015comprehensive}, the privacy preserving techniques are classified according to data distribution, data distortion, data mining algorithms, anonymization, data or rules hiding, and privacy protection. In the past decades, the problem of utility-oriented pattern mining (called utility mining or UPM for short) was proposed and has been extensively studied, including High Utility Itemset Mining (HUIM) \cite{yao2006mining,tseng2013efficient,gan2018survey}, High Utility Sequential Pattern Mining (HUSPM) \cite{ahmed2010novel,yin2012uspan,lan2014applying}, High Utility Episode Mining (HUEM) \cite{wu2013mining}, etc. Here, the term of ``\textit{utility}'' refers to the concept from \textit{utility theory} \cite{marshall2009principles}, that is different from the so-called utility (refers to availability of data) in previous PPDM framework and algorithms. Utility mining has been shown a powerful analytical ability to explore data in various real-life applications. Privacy-Preserving Utility Mining (PPUM) \cite{yeh2010hhuif,yun2015fast,lin2014reducing,lin2014efficiently,lin2016fast} which makes use of utility mining, has also become a critical issue in recent years. PPUM manages privacy concern by combining PPDM and utility-based pattern mining methods. Thus, we can say that PPUM is a sub-field of PPDM. The privacy and pattern utility are also two conflicting metrics in PPUM. Two approaches are commonly used, namely input privacy and output privacy. Input privacy changes the contents of database before conducting mining operations (i.e., perturbation \cite{bertino2008survey}, \textit{k}-anonymity \cite{sweeney2002k}, etc.). In the case of output privacy, contents of database are not changed, but rather data is made accessible to only intended people (i.e., secure multi–party computation \cite{lindell2000privacy,lindell2005secure}).  PPUM uses some technologies of utility mining and privacy preserving  as possible to preserve maximum utility. 

Several surveys present the state-of-the-art of privacy preserving techniques, mostly focusing on data aggregation, publishing and mining \cite{bertino2008survey,aggarwal2008general,vaidya2004privacy}. All of them are related to privacy preservation for data publishing and mining. However, none of them address the problem of privacy preserving for utility-based pattern mining. Yet, the concepts, utilization, categorization, and various characteristic of PPUM in terms of its strength and weakness are not methodically reviewed. To the best of our knowledge, this is the first survey to describe the basic concepts of PPUM, and summarize the current state-of-the-art privacy-preserving utility mining (PPUM) methods in detail. The main contributions of this paper are summarized as follows: 

\begin{itemize}
	\item We provide a detailed survey for existing privacy preserving utility mining (PPUM) techniques. Firstly, we present the background of utility mining, PPDM and PPUM, then introduce the related preliminaries and problem formulation of PPUM.
	 
	\item We summarize the literatures of utility mining and PPUM, and further present a general framework of existing PPUM techniques.  In particular, we also present and discuss the current state-of-the-art PPUM algorithms (e.g., each detailed algorithm for HUIM or HUSPM with privacy-preserving), as well as their advantages and deficiencies.
	
	\item Finally, we present some discussions of technical challenges and open directions for future research on PPUM.
\end{itemize}

The remainder of this survey is organized as follows. Section II introduces the related work of utility-based data mining and privacy preserving utility mining. Section III introduces the related preliminaries  and problem formulation of PPUM. Besides, the difference between utility mining and PPUM, and some key evaluation criteria for PPUM are reviewed in Section III. Section IV presents the current state-of-the-art PPUM algorithms, as well as their advantages and deficiencies. Section V discusses some challenges about PPUM and presents open issues for further research on PPUM. Section VI concludes this paper.

%% file: 2_relatedwork.tex
\section{Related work}

\subsection{Utility-Based Data Mining}

In the past decade, the problem of high-utility pattern mining (HUPM for short) \cite{yao2006mining,tseng2013efficient,gan2018survey} has been extensively studied. HUPM is different from other support-based pattern mining framework \cite{han2004mining}. Both the quantity and unit profit of objects/items are considered in HUPM to determine the importance of a high-utility pattern (HUP) rather than only its occurrence. Existing HUPM algorithms can be divided into the following categories \cite{gan2018survey}: Apriori-like approaches, tree-based approaches, utility-list-based approaches, and hybrid approaches. The early algorithms for HUPM are all the Apriori-like approaches using the generation-and-test mechanism, and the transaction-weighted utilization (TWU) model \cite{liu2005two} is widely adopted to keep the downward closure property for mining HUPs. The tree-based IHUP \cite{ahmed2009efficient}, HUP-tree algorithm \cite{linhup-tree}, UP-growth \cite{tseng2010up} and UP-growth+ \cite{tseng2013efficient} are all outperform the Apriori-like algorithms, while they perform worse than the utility-list-based algorithms, such as HUI-Miner \cite{liu2012mining}, d2HUP \cite{liu2012direct}, FHM \cite{fournier2014fhm}, and HUP-Miner \cite{krishnamoorthy2015pruning}. Recently, EFIM \cite{zida2017efim} was presented to efficiently mine the HUPs. Different from the above algorithms which aim at improving the mining efficiency, many studies focus on the mining effectiveness of HUPs. For example, mining high utility patterns from uncertain databases \cite{lin2016efficient} or temporal databases \cite{lin2015efficient,lin2017two}, HUPM with various discount strategies \cite{lin2016fast}, HUPM using multiple minimum utility thresholds \cite{lin2016efficient}, a condensed set of HUPs \cite{tseng2016efficient}, discriminative HUPs \cite{lin2017fdhup}, correlation utility patterns \cite{gan2018extracting}, and top-$k$ issue of HUPM \cite{tseng2016efficient}. Consider the time-ordered sequence data, some algorithms are developed to discover high-utility sequential patterns \cite{ahmed2010novel,yin2012uspan,lan2014applying}. At the same time, several studies about dynamic utility mining \cite{lin2015fast,2gan2018survey}, and HUPM from big data \cite{lin2015mining} have been introduced.  A comprehensive survey of utility-oriented pattern mining can be also referred to \cite{gan2018survey,2gan2018survey}.

\subsection{Privacy Preserving Utility Mining}

Knowledge discovery in databases (KDD, which is also called data mining) \cite{han2004mining} can find the hiding information and reveal the relationships among data. During this progress, it would also extract some sensitive information, which is not expected to be discovered. To solve the privacy problem, many privacy preserving data mining algorithms and technologies have been developed. Fayyad et al. first discussed the privacy problem in \cite{fayyad1996data}. They systematically discussed the privacy problem in KDD and highlighted some applications of KDD. In 2000, Agrawal et al. \cite{agrawal2000privacy} introduced a reconstruction procedure to accurately estimate the distribution of original data. Generally, the PPDM methodologies can be roughly divided into several categories, including pattern mining, classification, clustering, and so on  \cite{bertino2008survey,aggarwal2008general}. Note that this paper mainly focuses on privacy preserving with various patten mining tasks. Lindell et al.  \cite{lindell2000privacy} proposed a decision tree learning method based on ID3 to solve the problem of multi-party computation using private protocols \cite{lindell2005secure}). Many privacy preserving data mining methods are based on data distribution or perturbation, which does not consider the correlations between different dimensions. Aggarwal et al. \cite{aggarwal2004condensation} then proposed a condensation approach to solve this problem  by mapping the original database into a new anonymized database, and then condensing and grouping data, finally applying data mining algorithm. Vaidya et al. \cite{vaidya2004privacy} provided the details of privacy preserving data mining: why, how and when. In 2004, Verykios et al. \cite{verykios2004association} proposed three strategies and five algorithms for hiding association rules. These methods are all based on changing support or confidence or both to hide the sensitive rules. 
For frequent itemset hiding, Su et al. \cite{sun2005border} proposed a border-based approach which computes the border value for each sensitive frequent itemset, and then decides value to be decreased for PPDM. The KD-tree algorithm \cite{li2006tree} partitions original database to small sub-databases recursively, and then hides sensitive itemsets based on the average value of subsets. 

Since sensitive itemset usually contains more than one item, Li et al. proposed the MICF (maximum item conflict first) \cite{li2007micf} algorithm to hide sensitive patterns. MICF uses the maximum item which appears in sensitive itemsets as the target item in each step. In general, privacy preserving data mining considers both hiding effect and side effect. It is not realistic to hide sensitive itemsets without other side effects. Thus, how to find a balance between hiding effect and side effect is a critical issue. Wu et al. \cite{wu2007hiding} proposed a new PPDM method based on template which was built after classification. It generalizes the transactions and modification operation for PPDM. Moustakides et al. proposed a MaxMin \cite{moustakides2008maxmin} approach to hide the frequent patterns. Although these methods can efficiently hide the target of sensitive itemsets, but they do not consider the correlation among sensitive patterns. To address this problem, Hong et al. then introduced a TF-IDF based method named SIF-IDF \cite{hong2013using}. Lin et al. then proposed a series of methods \cite{lin2014reducing,lin2014efficiently}, which consider three side effects while hiding sensitive patterns.

%% file: 3_preliminaries.tex
\section{Preliminaries and Problem Formulation}

In this section, we first introduce some key preliminaries related to utility mining, and then describe the concept of PPUM, as well as some common evaluation criteria for PPUM. In particular, we highlight the relevance of utility mining to privacy preserving.

\subsection{Preliminaries of Utility Mining}

Let \textit{I} = \{\textit{i}$_{1}$, \textit{i}$_{2}$, $\ldots$, \textit{i$_{m}$}\} be a finite set of \textit{m} distinct items in a quantitative database \textit{D} = \{\textit{T}$_{1}$, \textit{T}$_{2}$, $\ldots$, \textit{T$_{n}$}\}, where each transaction  $ T_c = \{q(i_{1}, T_{c}), q(i_{2}, T_{c}), \ldots, q(i_{j}, T_{c})\} $  is a subset of \textit{I}, and has an unique identifier (\textit{tid}). Note that the $ q(i_{j}, T_{c}) $ is the quantity of each item $i_{j}$ in $T_c$. The unique profit $ pr(i_{j})$ is assigned to each item $i_{j} \in I$, which represents its importance (e.g., profit, interest, risk), as shown in a profit-table named \textit{ptable} = \{$ pr(i_{1}) $, $ pr(i_{2}) $, $\dots$, $ pr(i_{m}) $\}. An itemset with \textit{k} distinct items \{$ i_{1} $, $ i_{2} $, $\dots$, $ i_{k} $\} is called a \textit{k}-itemset. As a running example, Table \ref{db} shows a quantitative database containing 10 transactions, and we assume that the \textit{ptable} is defined as \textit{ptable} = \{\textit{pr}(\textit{A}): \$2, \textit{pr}(\textit{B}): \$6, \textit{pr}(\textit{C}): \$5, \textit{pr}(\textit{D}): \$1, \textit{pr}(\textit{E}): \$3, \textit{pr}(\textit{F}): \$8\}.

\begin{table}[!htbp]
	\setlength{\abovecaptionskip}{0pt}
	\setlength{\belowcaptionskip}{0pt} 	
	\caption{An example database}
	\label{db}
	\centering
	\begin{tabular}{|c|c|c}
		\hline
		\textbf{tid} & \textbf{Transaction (item, quantity)} \\ \hline
		$ T_{1} $ & 	\textit{A}:1, \textit{B}:9, \textit{C}:2, \textit{E}:5, \textit{F}:9  \\ \hline
		$ T_{2} $ & 	\textit{A}:7, \textit{C}:1, \textit{D}:7   \\ \hline
		$ T_{3} $ &	    \textit{B}:3, \textit{C}:1, \textit{D}:1, \textit{E}:1, \textit{F}:5  \\ \hline
		$ T_{4} $ &	    \textit{D}:5, \textit{E}:5  \\ \hline
		$ T_{5} $ &	    \textit{B}:1, \textit{E}:10, \textit{F}:7 \\ \hline
		$ T_{6} $ &	    \textit{C}:2, \textit{D}:2  \\ \hline
		$ T_{7} $ &	    \textit{A}:7, \textit{C}:5, \textit{D}:10, \textit{E}:10, \textit{F}:8  \\ \hline
		$ T_{8} $ &	    \textit{B}:4, \textit{C}:9, \textit{D}:9, \textit{E}:9  \\ \hline
		$ T_{9} $ &	    \textit{D}:7, \textit{E}:7, \textit{F}:5  \\ \hline
		$ T_{10} $	&	\textit{A}:2, \textit{B}:2, \textit{C}:7, \textit{D}:4, \textit{E}:5, \textit{F}:3  \\ \hline	
	\end{tabular}
	
\end{table}

\begin{definition}
	\label{def_1}
	\rm Given a quantitative database $D$, the utility of an item $ i_{j} $ in a transaction $ T_{c} $ is denoted as $u(i_{j}, T_{c}) $ and defined as $ u(i_{j}, T_{c})$ = $q(i_{j}, T_{c}) $ $\times $ $pr(i_{j}) $. The utility of an itemset \textit{X} in a transaction $ T_{c} $ is denoted as $ u(X, T_{c}) $ and defined as $ u(X, T_{c})$ = $\sum _{i_{j}\in X\wedge X\subseteq T_{c}}u(i_{j}, T_{c}) $. Thus, let  $ u(X) $ denote the utility of \textit{X} in \textit{D}, we have $ u(X) = \sum_{X\subseteq T_{c}\wedge T_{c}\in D} u(X, T_{c}) $.
\end{definition}

For example, the utility of (\textit{C}) in transaction $ T_{1} $ is calculated as $ u(C, T_{1}) = q(C, T_{1})\times pr(C) $ = 2 $ \times$ \$5 = \$10. And the utility of itemset $ (AC) $ in $ T_{1} $ is calculated as $ u(AC, T_{1}) $ = $ u(A, T_{1}) $ + $ u(C, T_{1}) $ = $ q(A, T_{1}) \times pr(A)$ + $ q(C, T_{1})\times pr(C) $ = $ 1 \times \$2 $ + $ 2 \times \$5 $ = \$12. Thus, the utility of $ (AC) $ is calculated as $ u(AC) $ = $ u(AC, T_{1}) $ + $ u(AC, T_{2}) $ + $ u(AC, T_{7}) $ + $ u(AC, T_{10}) $ = \$12 + \$19 + \$39 + \$39 = \$109.

\begin{definition}
	\label{def_2}
	\rm The transaction utility of a transaction $ T_{c} $ is denoted as $ tu(T_{c}) $ and defined as $ tu(T_{c}) = \sum_{i_{j}\in T_{c}}u(i_{j}, T_{c}) $, in which $j$ is the number of items in $T_{c}$. The total utility in \textit{D} is the sum of all transaction utilities and denoted as $TU$, which can be defined as $ TU=\sum _{T_{c}\in D}tu(T_{c}) $. Thus, an itemset \textit{X} in a database is said to be a high-utility itemset (HUI) if its total utility in the database is no less than the minimum utility threshold ($minutil$) multiplied by the $TU$, such as: $ HUI\leftarrow\{X|u(X)\geq minutil \times TU\} $.

\end{definition}

The above definitions are related to itemset-based data, and they have been extended to the sequence-based data \cite{yin2012uspan,lan2014applying}, which can be defined as follows.

\begin{definition}
	\rm The utility of an item ($ i_{j} $) in a \textit{q}-itemset \textit{v} is denoted as $ u(i_{j}, v) $, and defined as	$u(i_{j}, v) = q(i_{j}, v)\times pr(i_{j}),$ where $ q(i_{j}, v) $ is the quantity of ($ i_{j} $) in $ v $, and $ pr(i_{j}) $ is the profit of ($ i_{j} $). The utility of a \textit{q}-itemset $ v $ is denoted as $ u(v) $ and defined as	$u(v) = \sum_{i_{j}\in v}u(i_{j}, v)$. The utility of a \textit{q}-sequence $ s $ = $<$$v_{1}, v_{2}, \dots, v_{d}$$>$ is denoted as $ u(s) $ and defined as $u(s) = \sum_{v\in s}u(v)$.  A sequence $ s $ in a quantitative sequential database \textit{QSD} is said to be a \textit{high-utility sequential pattern} (HUSP) if its total utility is no less than the minimum threshold  multiplied by the $TU'$. That is $HUSP\gets\{s|u(s) \geq minutil \times TU'\}$, in which  $TU'$ is the total utility of \textit{QSD}. Consider the time-ordered sequences, high-utility sequential pattern mining (HUSPM) \cite{ahmed2010novel,yin2012uspan,lan2014applying} can discover more informative sequential patterns. This process is more complicated than the high-utility itemset mining \cite{yao2006mining,tseng2013efficient} and sequential pattern mining \cite{pei2004mining} since both the order and the utilities of sequences are considered together. 
\end{definition}

In this paper, the term ``HUPs" means each type of high-utility patterns, either HUIs or HUSPs.

\subsection{Privacy-Preserving Utility Mining}

\begin{figure}[hbtp]
	\centering
	\includegraphics[scale=0.3]{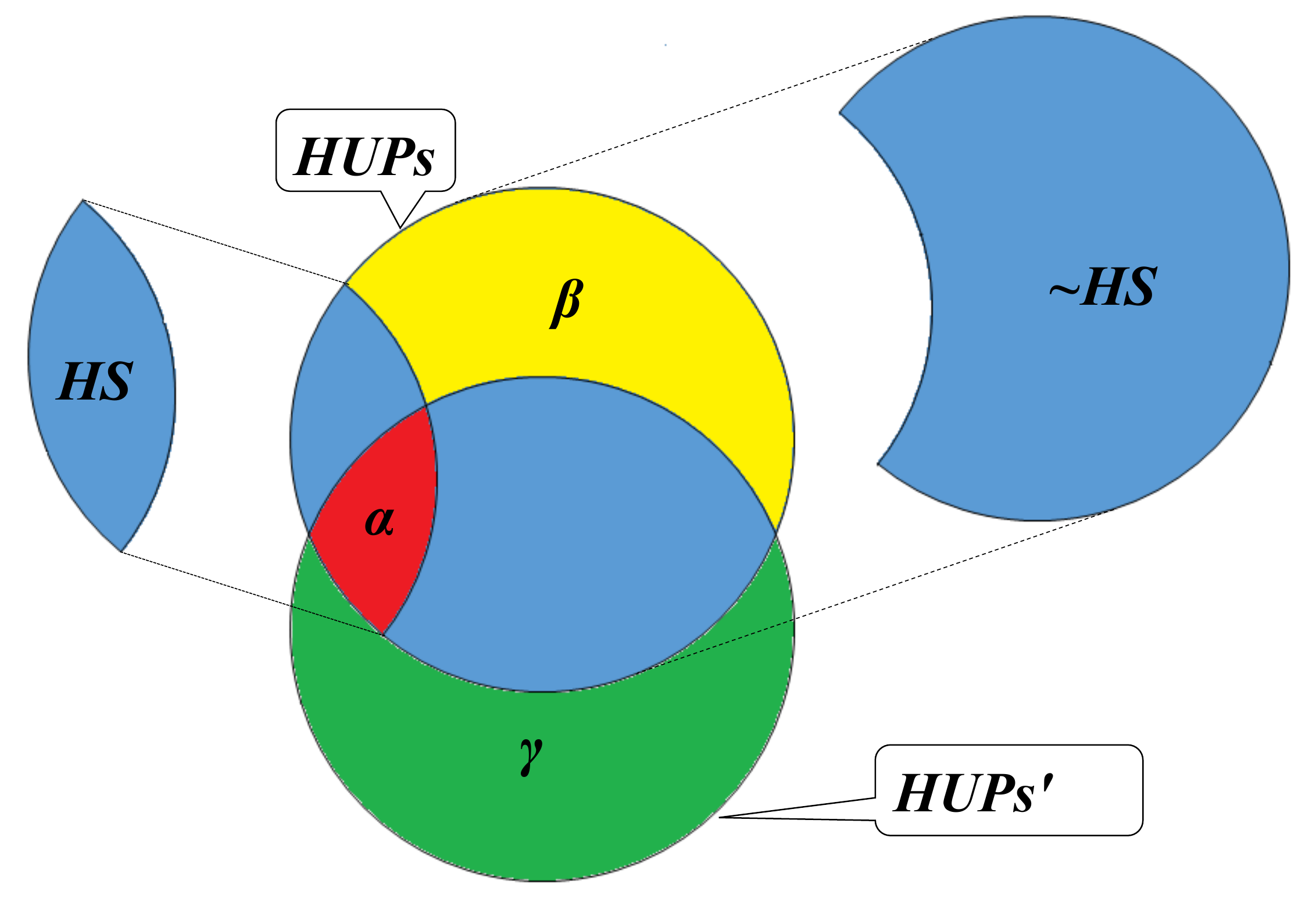}
	\captionsetup{justification=centering}
	\caption{The relationships between three side effects and the discovered HUPs (before and after sanitization) \cite{lin2017efficient}.}
	\label{fig:relationship}
\end{figure}

In general, some data would be lost when the original database is modified for privacy preserving. There is an important question that how to measure the sanitization cost. Bertino et al. \cite{bertino2005framework} proposed a framework to evaluate the performance of privacy preserving data mining. Three side effects in PPDM, including \textit{hiding failures} (HF), \textit{missing cost} (MC) and \textit{artificial cost} (AC) are presented. Most researchers  \cite{yeh2010hhuif,lin2014reducing,verykios2004association,li2007micf,wu2007hiding,hong2013using,lin2014ga} adopt these three side effects to measure the performance of their proposed algorithms. Hiding failure means some information has not been hidden completely after the sanitization process. The attacker may still be able to extract the sensitive information from the final sanitized database. Missing itemsets/rules mean some non-sensitive but large (also called frequent) itemsets/rules become not large anymore after sanitization. Thus, missing cost is the ratio of missing itemsets/rules, and lower missing cost is better for PPDM. Artificial itemsets/rules are those patterns which become large after the hiding process while they are small (also called non-frequent) in the original database. A high artificial cost may lead to a low accuracy since most artificial results are considered as the noise. Let $HS$ = \{$s_1$, $s_2$, $\dots$, $s_k$\} denote the set of sensitive HUPs (e.g., HUIs, HUSPs) to be hidden in a database $D$. The relationships between the side effects and mined patterns from the original database and sanitized patterns are shown in Fig. \ref{fig:relationship}. According to the definitions by Lin et al. \cite{lin2017efficient}, several criteria similar to PPDM \cite{wu2007hiding} are used for PPUM \cite{yeh2010hhuif}, and the details are defined as follows.

\begin{definition}
  \rm Let $\alpha $ (= $HF$) be the sensitive HUPs that the sanitization process failed to hide, that is the number of sensitive HUPs that still appears in the database after sanitization process. Formally, it is defined as:
	\begin{equation}
       \alpha = HS \cap HUPs',
    \end{equation} 	
 where $HS$ is the set of sensitive HUPs before the sanitization process, and HUPs is the set of high-utility patterns after the sanitization process.
\end{definition}

\begin{definition}
	\rm Let $\beta $ (= $MC$) be the missing HUPs, i.e., the HUPs that are non-sensitive but would be hidden after sanitization as: 
	\begin{equation}
         \beta = \sim HS - HUPs'.      
\end{equation}

\end{definition}

\begin{definition}
	\rm Let $\gamma $ (= $AC$) be the itemsets that were not HUPs before sanitization process but become HUPs in the sanitized database, that is the difference between HUPs' and HUPs as:
	\begin{equation}
        \gamma = HUPs' - HUPs,
    \end{equation}
where HUPs' is the set of HUPs obtained after the sanitization process.
\end{definition}

The relationships between these three side effects, and the discovered high-utility itemsets (before and after sanitization are respectively denoted as \textit{HUPs} and \textit{HUPs}') are illustrated in Fig. \ref{fig:relationship} \cite{lin2017efficient}. Based on the above concepts, the formal problem statement of PPUM studied in this work is defined below.

\begin{figure}[htbp]
	\centering  
	\includegraphics[trim=10 25 5 25,clip,scale=0.65]{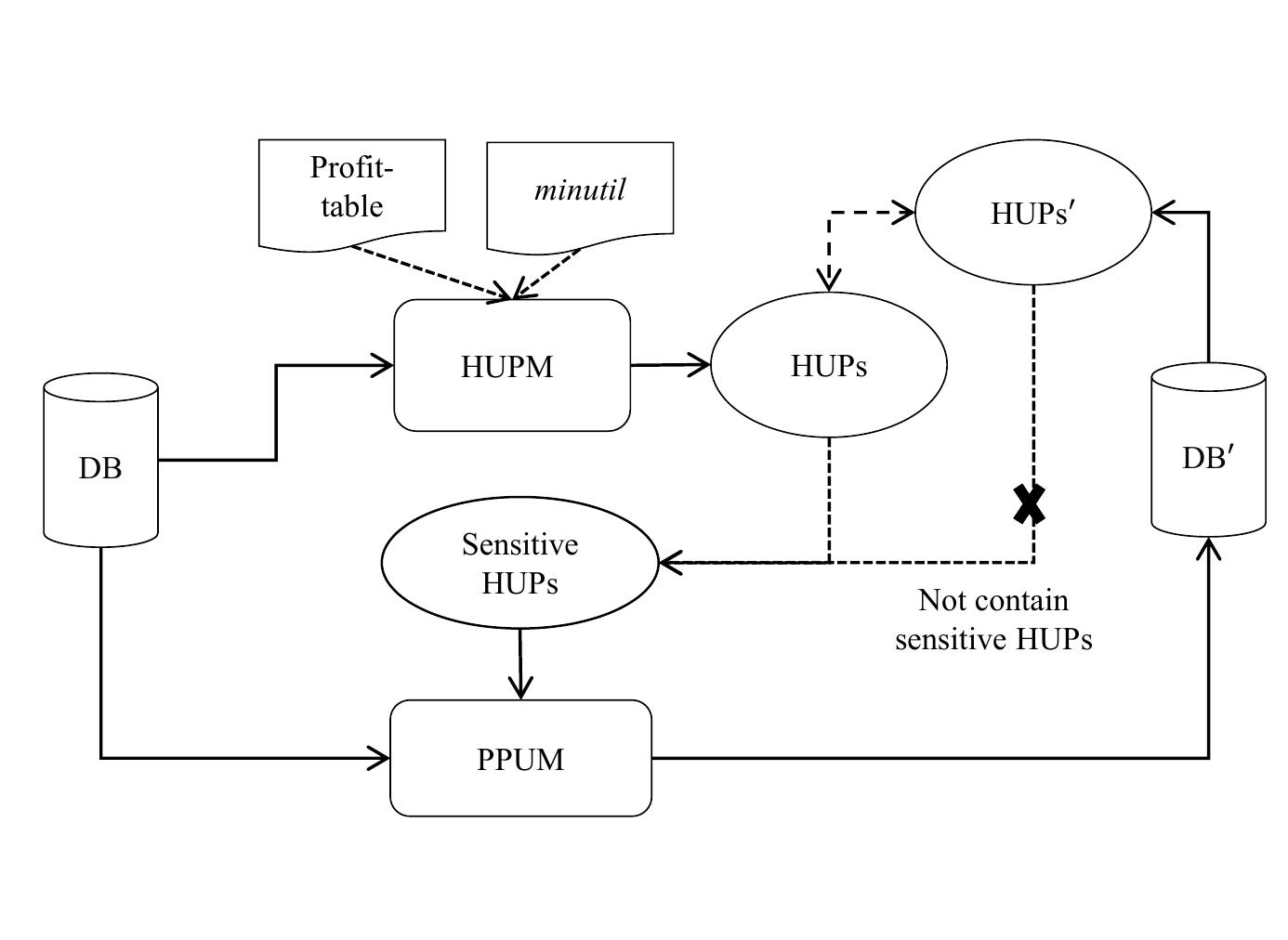}
	\captionsetup{justification=centering}
	\caption{Architecture of PPUM.}
	\label{fig:PPUM}	
\end{figure}

\textbf{Problem Statement}: Given a set of the sensitive high-utility patterns (HUPs, such as HUIs, HUSPs) to be hidden as $HS$ = \{$s_1$, $s_2$, $\dots$, $s_m$\}. One object of Privacy-Preserving Utility Mining (PPUM) is to completely hide the sensitive high-utility patterns as much as possible. Moreover, the designed algorithms of PPUM are to diminish the side effects as minimal as possible. Therefore, the goal of PPUM aims at finding an optimal solution for hiding as much as sensitive high-utility patterns as possible, while reducing and minimizing three side effects (\textit{HF}, \textit{MC} and \textit{AC}). Fig. \ref{fig:PPUM} shows the general architecture of PPUM.

% % % % % % % % % % % % % % %
%\vspace{-2.0em}
% % % % % % % % % % % % % % %

\begin{table*}[!htbp]
	\centering
	\scriptsize
	\caption{Comparison of hiding strategies.}
	\label{table_hidingStrategies}
	\newcommand{\tl}[1]{\multicolumn{1}{l}{#1}} %左对齐 
	\begin{tabular}{|c|l|l|l|l|} 
		\hline
		\multicolumn{1}{|c|}{\textbf{Name}} & \multicolumn{1}{|c|}{\textbf{Methodology (transaction and item)}} & \multicolumn{1}{|c|}{\textbf{Pros. and Cons.}} &  \multicolumn{1}{|c|}{\textbf{Year}} \\ \hline
		
		HHUIF \cite{yeh2010hhuif} & 	\multirow{2}{5cm}{Transaction: target item in the transaction. Item: the highest utility item.}  & \multirow{2}{6cm}{Straightforward, and performance is not optimized.}   & 2010  \\ 
		& &  &   \\ \hline
		
		MSICF \cite{yeh2010hhuif} &  \multirow{2}{5cm}{Transaction:  highest utility item in target transaction. Item: the highest occurrence item.}  & \multirow{2}{6cm}{It has good performance on condense database, but has high overlap sensitive itemsets.}  & 2010  \\  
		& &  &   \\ \hline

		FPUTT  \cite{yun2015fast}  & 	\multirow{2}{5cm}{Transaction: target item in the transaction. Item: the highest utility item.} & \multirow{2}{6cm}{Use tree structure to obtain better running time.} &  2014  \\
	 
		& &  &    \\ \hline

		GA-based insertion  \cite{lin2014reducing}  & 	\multirow{2}{5cm}{Insert appropriate transactions based on genetic algorithm.} & \multirow{2}{6cm}{Theoretical optimal solutions can be obtained but has to set the parameters, which might affect the final results.} &  2014  \\   
		& &  &    \\ \hline

		GA-based deletion  \cite{lin2014efficiently}  & 	\multirow{2}{5cm}{Delete appropriate transactions based on genetic algorithm.} & \multirow{2}{6cm}{Theoretical optimal solutions can be obtained but has to set the parameters, which might affect the final results.} & 2013  \\ 
		& &  &   \\ \hline

		MSU-MAU \cite{lin2016fast} & 	\multirow{2}{5cm}{Transaction: maximum sensitive utility. Item: minimum utility item.} & \multirow{2}{6cm}{Lower utility lost with better side effects.} &  2017  \\
		& &  &   \\ \hline
		
		MSU-MIU \cite{lin2016fast} & 	\multirow{2}{5cm}{Transaction: maximum sensitive utility. Item: minimum utility item.} & \multirow{2}{6cm}{Lower utility lost with better side effects.} &  2017  \\
		& &  &   \\ \hline

	\end{tabular}
\end{table*}

\subsection{Difference between PPUM and Utility Mining}

In general, utility mining mainly focuses on mining interesting patterns which have high utilities no less than the minimum utility threshold, and  discovering insight of the mined knowledge. Thus, utility mining cannot guarantee the protection of sensitive data. However, data with privacy is vitally important in many data analysis applications. Thus, the privacy problem also exists in high-utility pattern mining. In some real-world applications, the sensitive high-utility patterns (i.e., itemsets, sequences, episodes, etc.) \cite{gan2018survey} are usually needed to be hidden. Privacy-preserving utility mining (PPUM) is a critical issue, and it becomes an active research topic in recent years. Therefore, PPUM conducts data mining operations under the condition of preserving data privacy. If we emphasize data privacy, we may compromise the benefits of utility mining. PPUM towards to get balance between privacy and utility factors, which are two conflicting metrics. Thus, it is a non-trivial task to find the optimized solutions between the two conflicting metrics.

\subsection{The Hiding Strategies}
For privacy-preserving utility mining, both quantity and profit of each object (e.g., item, sequence, episode) are considered in sanitization process. The overall utility of a specific pattern in a database is defined as the sum of quantities multiplied by its unit profit. In general, there are two ways to decrease the utility value of a pattern: 1) directly decrease the quantity of a specific pattern, and 2) change the unit profit value to decrease the value of this pattern. In real-life situation, the profit would not be changed obviously. Thus, decreasing the occurred quantity of object is more reasonable and acceptable. Detailed comparison of the hiding strategies on PPUM are described in Table \ref{table_hidingStrategies}.

% % % % % % % % % % % % % % %
%\vspace{-2.0em}
\begin{figure*}[htbp]
	\centering %所给出的四个数字 分别代表了从左、下、右、上被截去的值
	\includegraphics[trim=5 20 18 10,clip,scale=0.32]{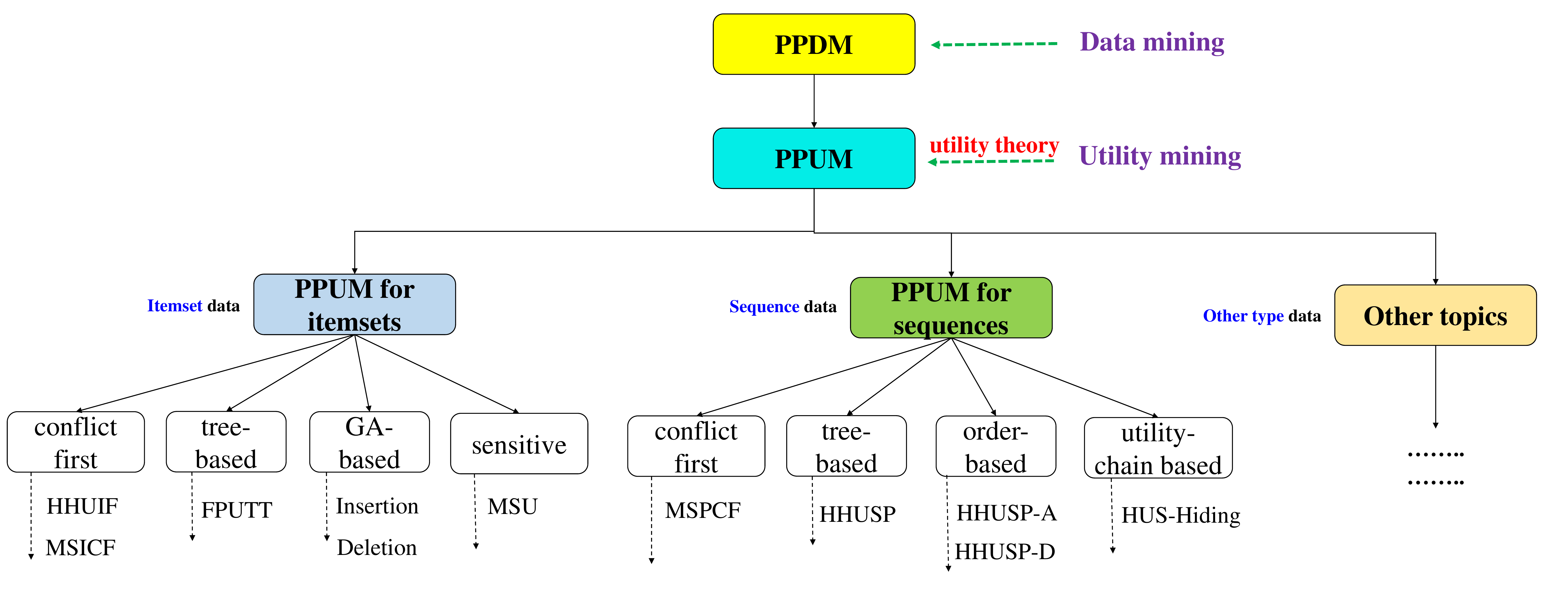}
	\captionsetup{justification=centering}
	\caption{Taxonomy of the existing PPUM algorithms.}
	\label{fig:reviewOfPPUM}	
\end{figure*}
% % % % % % % % % % % % % % %

\subsection{Evaluation Criteria for PPUM}

As mentioned before, up to now, many algorithms for PPUM \cite{yeh2010hhuif,lin2014reducing,verykios2004association,li2007micf,wu2007hiding,hong2013using,lin2014ga}, and most of them adopt three side effects \cite{bertino2005framework} (including hiding failure, missing cost, and artificial cost) to measure the performance of the proposed algorithms. The relationships between these side effects and mined patterns from the original database and sanitized one have shown in Fig. \ref{fig:relationship}.

High-utility pattern mining (HUPM) considers both quantity and profit for each item. Thus, privacy-preserving utility mining (PPUM) is different  compared with PPDM in evaluation criteria. To address the problem of PPUM, Lin et al. \cite{lin2016fast} proposed another three criteria for performance evaluation namely \textit{database structure similarity} (DSS), \textit{database utility similarity} (DUS), and \textit{itemset utility similarity} (IUS). Rajalaxmi and Natarajan also proposed a similar concept namely \textit{utility difference} \cite{rajalaxmi2012effective} with DUS. Details of the related definitions are shown as follows.

\begin{definition}
	\label{def_5}
	\rm Let $D$ denote the database, \textit{I} = \{\textit{i}$_{1}$, \textit{i}$_{2}$, $\ldots$, \textit{i$_{m}$}\} be a finite set of \textit{m} distinct items in $D$. A transaction $T_i$ is a subset of $I$, $T_i\in I$. That is defined as: $TP(T_i)$ = $\{(v_1, v_2, \dots, v_k)|v_{j} = 1, v_{j}\in T_{i}; v_{j} = 0, v_{j}\notin T_i; 1 \leq j \leq k\}$, and $ TP = \{TP(T_i) | T_{i} \in D; TP(T_i) \neq TP(T_j), i \neq j\}$.
\end{definition}

The transaction pattern is a set of items. For example, there are six items $I$ = $\{A, B, C, D, E, F\}$ in Table \ref{db}, transaction $T_2$ = \{$A$:7, $C$:1, $D$:7\}, $TP(T_2)$ = \{1, 0, 1, 1, 0, 0\}. Thus, \textit{transaction pattern} is related to the occurred items in this transaction. Table \ref{db} shows the correspondence relation between \textit{transaction} and \textit{transaction pattern}. Usually, \textit{transaction pattern} is called \textit{pattern}. $TP(i)$ can be also referred as $TP(T_i)$ for simplicity.

\begin{definition}
	\label{def_6}
	\rm Let $TP$ denote the original transaction pattern derived from $D$ and $TP'$ denote the new set derived from the perturbed database $D'$. The \textit{database structure similarity} (DSS) \cite{lin2016fast} is defined as:
	\begin{equation}
	DSS = \dfrac{\sum\limits_{i=1}^{n}size(TP_i) \times size(TP'_i)}{\sqrt{\sum\limits_{i=1}^{n}size(TP_i)^2} \times \sqrt{\sum\limits_{i=1}^{n}size(TP'_i)^2}},
	\end{equation}
where $n$ = $max\{|TP|, |TP'|\}$, $size(TP_i)$  equals to the number of transactions such that $TP(j)$ = $TP_i$, 0  $ \leq j \leq |D|$.
\end{definition}

\begin{definition}
	\label{def_7}
	\rm Let $D$ and $D'$ respectively denote the original database and the sanitized database. The loss utility between the original database and the sanitized database is measured by \textit{Database Utility Similarity} (DUS) \cite{lin2016fast}, which can be defined as:
	\begin{equation}
	DUS = \dfrac{\sum\limits_{T_c \in D'}tu(T_c)}{\sum\limits_{T_c \in D}tu(T_c)}.
	\end{equation}
\end{definition}

Note that the concept of \textit{utility integrity} (UI) which was proposed in PPUMGAT \cite{lin2017efficient}, in fact, is the same as the DUS concept. The \textit{utility integrity} evaluates the difference in terms of total utility before and after sanitization.

\begin{definition}
	\label{def_8}
	\rm Let $HUPs^D$ and $HUPs^{D'}$ denote the discovered high-utility patterns (e.g., HUIs, HUSPs) from the original database $D$ and the sanitized database $D'$, respectively. The loss utilities of the discovered HUPs before and after sanitization is denoted as \textit{Itemset Utility Similarity} (IUS) \cite{lin2016fast}, which can be defined as:
	\begin{equation}
	IUS = \dfrac{\sum\limits_{X \in HUPs^{D'}}u(X)}{\sum\limits_{X \in HUPs^D}u(X)}.
	\end{equation}
\end{definition}

%% file: 4_survey.tex
\section{State-of-the-Art Algorithms for PPUM}

\subsection{Overview of PPUM}

Firstly, an overview of the developed algorithms in PPUM is briefly provided here. Fig. \ref{fig:reviewOfPPUM} presents a detailed taxonomy of the existing PPUM algorithms. It can be seen that most of algorithms for PPUM deal with itemset-based or sequence-based data to address the problem of hiding sensitive patterns in utility mining.

\subsection{HHUIF and MSICF}

The main idea of HHUIF (hiding high utility item first algorithm) \cite{yeh2010hhuif} is to hide the sensitive HUIs one-by-one by recursively decreasing the quantity of the target items which having the highest utility value. For a specific sensitive HUI $si$, it projects the related transactions $D|_{si}$ and then selects the item $i \in SI$ which has the highest utility. The transaction where this target item $i$ located is called the target transaction and denoted as $T|i$. Then two strategies are used to modify the target item $i$. A difference value \textit{diff} is computed for each sensitive HUI. The \textit{diff} means how much we need to decrease to hide current sensitive HUI. For each step, if the \textit{diff} is larger than the utility of target item $i$, the item $i$ is then removed from the target transaction $T|i$, or the $ \lceil diff/u(i) \rceil $ of the quantity from $T|i$ is decreased. It repeats the hiding steps until \textit{diff} $\leq $ 0. The pseudo-code of HHUIF is shown in Algorithm \ref{AlgorithmOfHHUIF}.

MSICF (maximum sensitive itemsets conflict first algorithm) \cite{yeh2010hhuif} selects the item which has the maximum conflict count among items in the sensitive itemsets as the target item. MSICF is similar with HHUIF in database (transaction) modification. The difference between them is the way of selecting target items and target transactions. MSICF first calculates the conflict count for each sensitive item which is contained by sensitive itemsets. Then MSICF sorts these sensitive items in descending order and hides them one-by-one based on the sorted sensitive items. An item which has the highest conflict count means it is contained by more sensitive items. Thus, modify this item may affect more sensitive itemsets. In each step, it select the transaction which has the highest utility value of target item. MSICF repeats these processes until all sensitive itemsets have been hidden. Both HHUIF and MSICF are simple, however, their performance are not good enough and have the problem of side effects.

%%%%%%%%%%%%%%%%%   Algorithm HHUIF    %%%%%%%%%%%%%%%%%%%
%\floatname{algorithm}{Alg. HHUIF}%自己命名首行开头显示内容
\renewcommand{\algorithmicrequire}{\textbf{Input:}}%Input
\renewcommand{\algorithmicensure}{\textbf{Output:}}%Output
\begin{algorithm}
	\caption{HHUIF algorithm}\label{AlgorithmOfHHUIF}
	
	\begin{algorithmic}[1]		
		%输入Input
		\Require \textit{D}, the original database; \textit{SI}, the sensitive HUIs; \textit{minutil}, the minimum utility threshold.
		%输入Output
		\Ensure $D'$, a sanitized database without sensitive HUIs.

		\For{ each $si \in {SI}$} 
		\State \textit{diff} = \textit{u(si)} - \textit{minutil};\
		\State $D|_{si}\leftarrow$ projected database of \textit{si};\
		\While {$ diff \geq 0$)}
		\State $node(i_t, T_t)$ $\leftarrow$ \{$node(i, T)|max(u(i, T))$, $T \in SI-table(si)$, $i\in si$\};\
		
		\State $$ q(i_t, T_t)=\left\{
		\begin{aligned}
		0, u(i_t, T_t) < diff \\
		q(i_t, T_t) - \lceil diff/u(i_t)  \rceil,  u(i_t, T_t) \geq diff
		\end{aligned}
		\right.
		$$
		
		\State $$ diff =\left\{
		\begin{aligned}
		0, u(i_t, T_t) > diff \\
		diff - u(si, T_t), u(i_t, T_t) \leq diff
		\end{aligned}
		\right.
		$$

		\EndWhile
   \State \textbf{end while}
		
		\EndFor
  \State \textbf{end for}		
	\end{algorithmic}
\end{algorithm}
%%%%%%%%%%%%%%%%%   Algorithm 2   %%%%%%%%%%%%%%%%%%%

\subsection{FPUTT}

The FPUTT (fast perturbation algorithm using tree  and table structures) \cite{yun2015fast} is developed to hide sensitive itemsets, and two structures are used to speed up the perturbation process. It can achieve the perturbation process with only three times of database scans. FPUTT first scans database to build the head-table. Note that the head-table only contains the items which are contained by sensitive itemsets. Assume that the sensitive items in Table \ref{db} are \{$B, C, D, E, F$\}. Thus, \{$T_1, T_3, T_5, T_7, T_8, T_{10}$\} contain sensitive itemsets, and the elements in head-table are \{$B$:5, $C$:5, $D$:4, $E$:6, $F$:5\}. Then they are sorted in descending order of count values as \{$E$:6, $B$:5, $C$:5, $F$:5, $D$:4\}.

% % % % % % % % % % % % % % %
%\vspace{-2.0em}
\begin{figure}[htbp]
	\centering %所给出的四个数字 分别代表了从左、下、右、上被截去的值
	\includegraphics[trim=0 0 5 5,clip,scale=0.39]{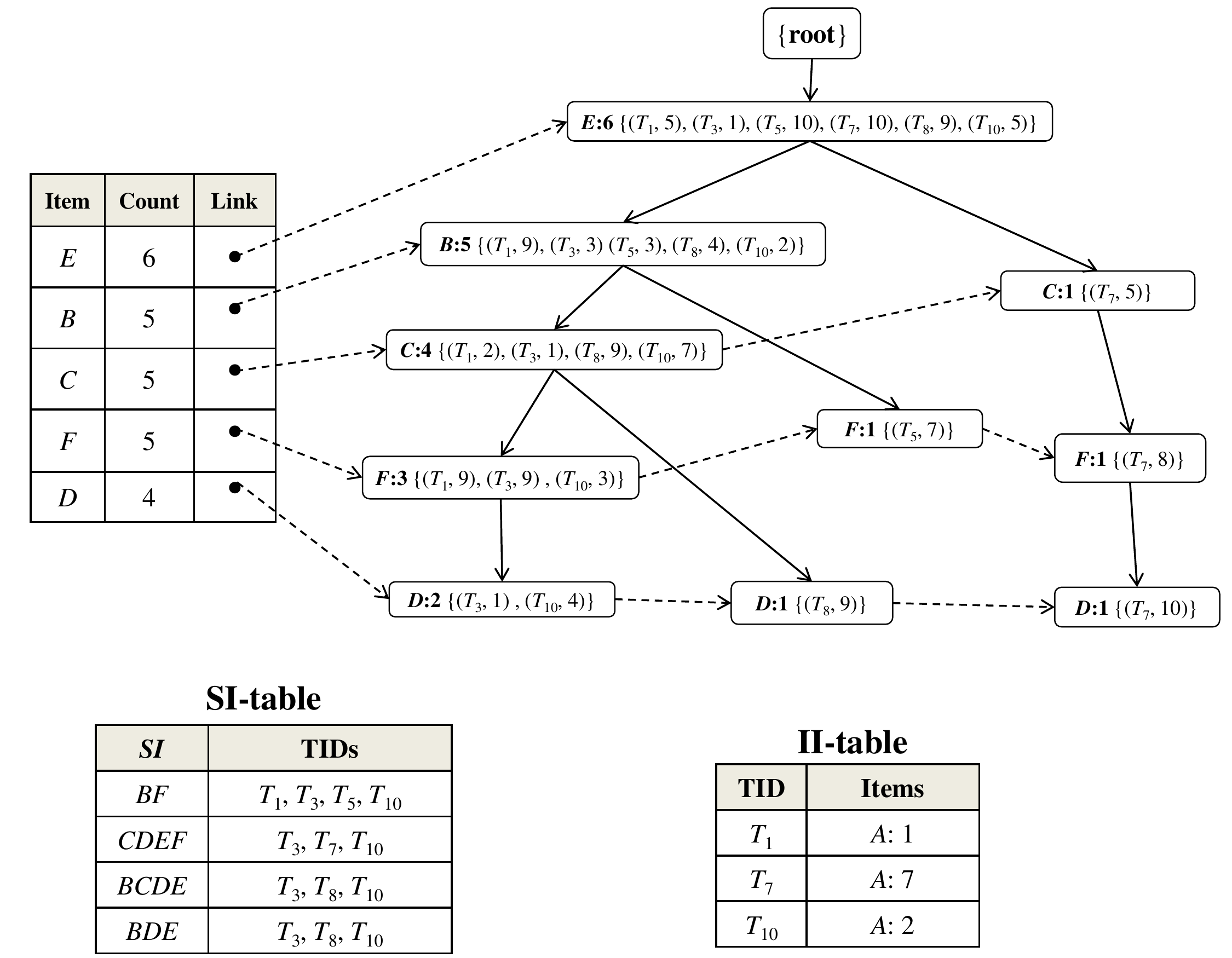}
	\captionsetup{justification=centering}
	\caption{The built FPUTT-tree of the example database.}
	\label{fig:FPUTT_tree}	
\end{figure}
% % % % % % % % % % % % % % %

Next, FPUTT rescans the database once to build FPUTT-tree based on head-table. In each insertion step, it needs to update both SI-table and II-table. The built tree structure of the given example is shown in Fig. \ref{fig:FPUTT_tree}. After building the complete FPUTT-tree, it then hides the sensitive itemsets one-by-one. Based on SI-table and II-table, the target items can be quickly found and modified. Although FPUTT is faster than previous algorithms, FPUTT-tree may not be compact and still occupy a large memory space. Its mining performance is related to the number of conditional trees and their construction/traversal cost. One of the performance bottlenecks is the generation of a huge number of conditional trees, which has highly computational and memory usage cost.

\subsection{MSU-MAU and MSU-MIU}

Recently, Lin et al. proposed two algorithms, called Maximum Sensitive Utility-MAximum item Utility (MSU-MAU) \cite{lin2016fast} and Maximum Sensitive Utility-MInimum item Utility (MSU-MIU) \cite{lin2016fast}, to minimize the side effects of the sanitization process for hiding sensitive HUIs. They use the projection mechanism and the concept of maximum utility to delete items or to decrease their quantities, thus reducing the utilities of sensitive HUIs in projected transactions. 

The MSU (maximum sensitive utility) algorithm can efficiently hide the sensitive HUIs. It is quite different from the previous methods, which selects the item having the maximum utility value as the target item. Based on the target items, it determines the target transactions. MSU first selects the target transaction in a greedy-based mechanism. In each step, it selects the transaction which has the maximum sensitive utility value and then finds out the item having the minimum utility value in the target transaction. The reason why selecting the minimum item is that there is same effect on the sensitive itemset whatever this item is deleted. Thus, there is little loss of database utility when selecting the minimum utility item and fewer modification in the original database.

For each sensitive HUI, the difference utility between the processed sensitive HUI and the minimum utility threshold is determined to check whether the related procedure is required to be processed. If the difference value is less than 0, it indicates that this HUI has been hidden; otherwise, it is needed to be hidden. Then MSU-MIU scans the projection database to find the target transactions and items. In each step, the transaction which has the maximum sensitive utility could be identified, and the item which has the minimum utility value in the target transaction could be selected. Finally, it returns the sanitized database after the hiding process is completed.

\subsection{Heuristic Hiding Algorithms}

Traditional methods of PPUM focus on how to find the best item to delete or to decrease its quantity. It is a NP-hard problem to find the best solution for hiding sensitive HUIs. Up to now, some artificial intelligence algorithms (e.g., genetic algorithm \cite{davis1991handbook}, particle swarm optimization \cite{kennedy2011particle}) are used to find the approximate optimal solutions. These algorithms can find a converge solution through iteration specific steps. Based on some properties of heuristic algorithms, two GA-based methods were proposed to hide sensitive HUIs \cite{lin2014reducing,lin2014efficiently}. The main idea of this method to hide sensitive HUIs is encoding the chromosome with transaction identity. A fitness function was also defined for evaluating the fitness value of the individual item. The fitness function considers all the three common side effects, and is defined below:
\begin{equation}
fitness(i) = w_1 \times \alpha + w_2 \times \beta + w_3 \times \gamma,
\end{equation}
in which $\alpha$, $ \beta $ and $ \gamma $ are the hiding failure, missing cost and artificial cost, respectively. $w_i$ is the weight value of each side effect. 

Recently, Lin et al. first proposed an optimisation approach namely PPUMGAT \cite{lin2017efficient} to hide sensitive HUIs based on a GA and the operation of transaction deletion. The purpose of PPUM is to find appropriate transactions to be deleted for hiding the sensitive HUIs, while minimizing the three side effects (hiding failures, missing cost, and artificial cost \cite{bertino2005framework}). An improved pre-large \cite{hong2001new} concept was also extended to speed up the evolution process. Based on the designed pre-large concept, multiple database scans can be avoided and runtime is thus decreased. The conducted experiments have shown that  PPUMGAT+ can successfully hide all sensitive HUIs, while maintaining high utility of database.

\subsection{PPUM for Sequence Data}

Up to now, many algorithms have been designed for hiding high-utility itemsets  \cite{yeh2010hhuif,yun2015fast,lin2014reducing,lin2014efficiently,lin2016fast}. However, it is a non-trivial task to extend these algorithms for handing sequence data. Due to the time order in sequence data, high-utility sequential pattern mining (HUSPM) \cite{ahmed2010novel,yin2012uspan,lan2014applying} is more difficult and complex than HUIM. Few algorithms have been proposed to hide frequent sequential patterns (FSP) in sequence databases. For example, the DBSH and SBSH \cite{gkoulalas2011revisiting} algorithms are developed to hide sequential patterns. However, they are not suitable for handling sequence databases with utility information. In fact, HUSPM is more difficult and challenging than mining FSP and hiding FSP. The reason is that the utility of sequence is neither monotonic nor anti-monotonic \cite{ahmed2010novel,yin2012uspan,lan2014applying}. For example, a HUSP may have a lower, equal or higher utility than any its sub-sequence. In order to hide high-utility sequential patterns (HUSPs) in sequence data, several related algorithms are further developed, such as HHUSP \cite{dinh2015novel}, MSPCF \cite{dinh2015novel}, HHUSP-A \cite{quang2016approach}, HHUSP-D \cite{quang2016approach} and HUS-Hiding \cite{le2018efficient}, as shown in Fig. \ref{fig:reviewOfPPUM}. Details of each algorithm are described below. 

$\bullet$  \textbf{HHUSP and MSPCF}. Inspired by the itemset-based HHUIF and MSICF algorithms \cite{yeh2010hhuif}, Dinh et al. first proposed two algorithms named HHUSP (Hiding High Utility Sequential Patterns) and MSPCF (Maximum Sequential Patterns Conflict First) to hide all HUSPs. Both of HHUSP and MSPCF \cite{dinh2015novel} use the  USpan \cite{yin2012uspan} algorithm in the first step to mine all HUSPs. Experimental results show that HHUSP run faster than MSPCF. However, the time of computing and memory usage of two algorithms are similar to that of USpan \cite{yin2012uspan} which may easily suffer from the computation problem. Besides, they do not consider  about the difference between the original database and the sanitized database after hiding phase.

$\bullet$  \textbf{HHUSP-A and HHUSP-D}. To address the shortcoming of HHUSP and MSPCF, Quang et al. then proposed two algorithms respectively called HHUSP-A \cite{quang2016approach} and HHUSP-D \cite{quang2016approach}. The HHUSP-A uses the ascending order of utility of HUSP, while the HHUSP-D relies on the descending order to improve the performance of HHUSP by decreasing execution time and missing cost. The main ideas of these four algorithms for hiding HUSPs had been illustrated in \cite{le2018efficient}. Their general process for hiding HUSP is composed of three steps: 1) Mining step: uses a HUSPM algorithm to mine all HUSPs form the original sequential database; 2) Sorting step: sorts the set of derived HUSPs using a specific order; and 3) Hiding step: modifies the original database and returns the final sanitized database. However, these three steps are usually time consuming in practice, especially when dealing with large-scale databases or a low minimum utility threshold.

$\bullet$  \textbf{MHHUSP and HUS-Hiding}. To further improve the hiding efficiency, the MHHUSP algorithm \cite{quang2016mhhusp} was proposed. Different from the previous algorithms using three steps to finish the hiding task, MHHUSP combines mining and hiding HUSP in a same process. However, despite this improvement, hiding HUSPs remains a very time and memory consuming process. Recently, Le et al. presented a novel algorithm named HUS-Hiding \cite{le2018efficient} which relies on a novel structure to enhance the sanitization process. Experimental results show HUS-Hiding  has a better performance than the previous PPUM algorithms for sequence data.

%% file: 5_opportunities.tex
%%%%%%%%%%%%%%%%%%%%%%%%%%%%%%%%%%%%%%%%%%%%%%%%%%%%%%%
%%%%%%%%%%%%%%%%%   Opportunities  %%%%%%%%%%%%%%%%%%%%%%%%
%%%%%%%%%%%%%%%%%%%%%%%%%%%%%%%%%%%%%%%%%%%%%%%%%%%%%%%

\section{Open Challenges and Opportunities}

In above section, we have reviewed the state-of-the-art algorithms of PPUM. Up to now, many utility-based mining framework and algorithms have been proposed. How can PPUM be able to take advantage of the advanced techniques of the existing mining framework and algorithms may lead to some opportunities. In particular, recent events and development have led to increasing interestingness by applying data mining techniques toward privacy-related problems. This also leads to the interestingness of technical challenges at the intersection of privacy, security and data mining. Here, we discuss several open challenges and opportunities for privacy-preserving utility mining.

$\bullet$  \textbf{\textit{What is privacy? What is private data? How to measure privacy?}} These are the fundamental problems for PPUM. Although several algorithms of PPDM have been extensively studied, there is still no universally standard and definition of privacy, i.e., personal privacy has different definitions. How to define an unified or specialized privacy concept may be important and interesting in different real-world applications. Similarly, the problem of what is private data may also lead to some opportunities. For example, whether the data value protection or data pattern protection is required for the specific applications? Furthermore, how to measure privacy in PPUM has potential technical challenges. Can we quantify privacy? How do we consider that privacy is preserved?

$\bullet$  \textbf{\textit{Application-driven algorithms}}. Up to now, most algorithms for PPUM have been developed to improve the evaluation of utility mining and efficiency of mining process. The domain application and effectiveness of the algorithms for PPUM is also very important. In general, the application-driven algorithms with some particular features of utility patterns reflect real-life problems of different applications in various fields \cite{gan2018survey}. This is a clearly opportunity to address these applications while using different PPUM techniques with  basic privacy and utility theory. Consider the information granulation in data science, how to address the PPUM issues related to Granular computing \cite{yao2016triarchic} and its applications is interesting. Privacy-preserving utility mining guided by domain knowledge also provides some opportunities.

$\bullet$  \textbf{\textit{Deal with complex data}}. With dramatic increasing of digital data, the ubiquitous data in the big data era is highly complex. Subsequently, many data mining techniques are developed in order to work with these complex data, such as ``structured data''\footnote{\url{https://en.wikipedia.org/wiki/Structure_mining}}, ``unstructured data''\footnote{\url{https://en.wikipedia.org/wiki/Unstructured_data}}, and ``semi-structured data''\footnote{\url{https://en.wikipedia.org/wiki/Semi-structured_data}}. Great advances in mining technology bring many benefits to society. At the same time, the growing privacy is also concerned especially in the areas of data mining and big data analytics. Privacy is a broad topic, and it encompasses a variety of issues in many different types of data. Most existing PPUM algorithms focus on itemset-based data, but few of them address the sequence data, even the graph data, stream data, distributed  and big data \cite{gan2017data}. There are some theoretical challenges for PPUM while dealing with these complex data. % in high dimensionality.

$\bullet$  \textbf{\textit{Developing more efficient algorithms}}. Although many utility mining framework and approaches have been proposed, some of these advanced techniques have not been utilized in the current PPUM algorithms. As shown before, most PPUM algorithms are computationally expensive in terms of execution time and memory cost. This may be a crucial problem while dealing with dense databases, databases containing long transactions, or a low minimum utility threshold chosen by the user. Although current PPUM algorithms are much efficient than previous algorithms, there is still room for improvement.

$\bullet$  \textbf{\textit{Flexible algorithms for PPUM}}. Another open challenge for PPUM is that how to deal with dynamic data but not the static data. All the existing PPUM algorithms are the batch models which are designed for handling static data. In some real-world applications, the processed data is dynamic changed overtime \cite{ahmed2009efficient,2gan2018survey}. If the batch model of PPUM is applied to this dynamic data, the privacy preserving problem would not be successfully solved. In other words, the sensitive utility patterns and information would not be protected while some records/data are added, deleted or modified.  Specifically, online model and dynamic model are, to a great extension, more complicated and challenging than the static model. Developing a flexible, interactive and adaptive model for privacy-preserving utility mining  leads to important future challenges.

\section{Conclusion}
\label{sec:conclusion}

Privacy-preserving utility mining (PPUM) is a relatively new area of research compared to the more established research areas of utility mining and privacy-preserving data mining. In this paper, we provide a comprehensive review of the current state-of-the-art PPUM algorithms, including the preliminaries of utility mining and PPUM, evaluation criteria for PPUM, details of existing PPUM algorithms (e.g., techniques, advantages and disadvantages). Finally, we highlight some important open challenges and opportunities of this topic that need to be further developed in the future. %We believe by reflecting upon the challenges described in this paper, we can take more concrete steps to achieve more flexible, interactive and adaptive PPUM models in real-world applications.

\section*{Acknowledgment}

This research was partially supported by the National Natural Science Foundation of China (NSFC) under grant No. 61503092, by the Shenzhen Technical Project under JCYJ20170307151733005 and KQJSCX20170726103424709. Specifically, Wensheng Gan was supported by CSC (China Scholarship Council) Program during the study at University of Illinois at Chicago, IL, USA.